\documentclass{ws-procs975x65}

\begin{document}

\markboth{Francisco S. N. Lobo}{Stable dark energy stars: An
alternative to black holes?}

\wstoc{Stable dark energy stars: An alternative to black
holes?}{Francisco S. N. Lobo}

\title{STABLE DARK ENERGY STARS: \\AN ALTERNATIVE TO BLACK HOLES?}

\author{FRANCISCO S. N. LOBO}

\address{Centro de Astronomia
e Astrof\'{\i}sica da Universidade de Lisboa,\\
Campo Grande, Ed. C8 1749-016 Lisboa, Portugal\\
\email{flobo@cosmo.fis.fc.ul.pt}}

\begin{abstract}

In this work, a generalization of the Mazur-Mottola gravastar
model is explored, by considering a matching of an interior
solution governed by the dark energy equation of state,
$\omega\equiv p/ \rho<-1/3$, to an exterior Schwarzschild vacuum
solution at a junction interface, situated near to where the event
horizon is expected to form. The motivation for implementing this
generalization arises from the fact that recent observations have
confirmed an accelerated cosmic expansion, for which dark energy
is a possible candidate.

\end{abstract}


\bodymatter

\section*{}

Although evidence for the existence of black holes is very
convincing, a certain amount of scepticism regarding the physical
reality of singularities and event horizons is still encountered.
In part, due to this scepticism, an alternative picture for the
final state of gravitational collapse has emerged, where an
interior compact object is matched to an exterior Schwarzschild
vacuum spacetime, at or near where the event horizon is expected
to form. Therefore, these alternative models do not possess a
singularity at the origin and have no event horizon, as its rigid
surface is located at a radius slightly greater than the
Schwarzschild radius. In particular, the gravastar ({\it
grav}itational {\it va}cuum {\it star}) picture, proposed by Mazur
and Mottola \cite{Mazur}, has an effective phase transition
at/near where the event horizon is expected to form, and the
interior is replaced by a de Sitter condensate.
In this work, a generalization of the gravastar picture is
explored, by considering a matching of an interior solution
governed by the dark energy equation of state, $\omega\equiv p/
\rho<-1/3$, to an exterior Schwarzschild vacuum solution at a
junction interface~\cite{darkstar2}. This new emerging picture
consisting of a compact object resembling ordinary spacetime, in
which the vacuum energy is much larger than the cosmological
vacuum energy, shall be denoted as a ``dark energy star''
\cite{Chapline}. We emphasize that the motivation for implementing
this generalization arises from the fact that recent observations
have confirmed an accelerated expansion of the Universe, for which
dark energy is a possible candidate. The dynamical stability of
the transition layer of these dark energy stars to linearized
spherically symmetric radial perturbations about static
equilibrium solutions was further explored~\cite{darkstar2}.

Consider the interior spacetime, without a loss of generality,
given by the following metric, in curvature coordinates
\begin{eqnarray}
ds^2&=&-e^{2\Phi(r)}\,dt^2+\frac{dr^2}{1- 2m(r)/r}
    +r^2 \,(d\theta ^2+\sin ^2{\theta} \, d\phi ^2) \label{metric}
\,,
\end{eqnarray}
where $\Phi(r)$ and $m(r)$ are arbitrary functions of the radial
coordinate, $r$. The function $m(r)$ is the quasi-local mass, and
is denoted as the mass function.

The Einstein field equation, $G_{\mu\nu}=8\pi T_{\mu\nu}$ provides
the following relationships
\begin{eqnarray}
m'&=&4\pi r^2 \rho  \,,  \qquad \Phi'=\frac{m+4\pi
r^3 p_r}{r(r-2m)} \label{Phi}\,,\\
p_r'&=&-\frac{(\rho+p_r)(m+4\pi r^3 p_r)}{r(r-2m)}
+\frac{2}{r}(p_t-p_r)\label{anisotTOV}\,,
\end{eqnarray}
where $'=d/dr$. $\rho(r)$ is the energy density, $p_r(r)$ is the
radial pressure, and $p_t(r)$ is the tangential pressure. Equation
(\ref{anisotTOV}) corresponds to the anisotropic pressure
Tolman-Oppenheimer-Volkoff (TOV) equation. The factor $\Phi'(r)$
may be considered the ``gravity profile'' as it is related to the
locally measured acceleration due to gravity, through the
following relationship: ${\cal A}=\sqrt{1-2m(r)/r}\,\Phi'(r)$. The
convention used is that $\Phi'(r)$ is positive for an inwardly
gravitational attraction, and negative for an outward
gravitational repulsion.

Now, using the dark energy equation of state, $p_r=\omega \rho$,
and taking into account Eqs. (\ref{Phi}), we have the following
relationship
\begin{equation}
\Phi'(r)=\frac{m+\omega rm'}{r\,\left(r-2m \right)} \,.
            \label{EOScondition}
\end{equation}
One now has at hand four equations, namely, the field Eqs.
(\ref{Phi})-(\ref{anisotTOV}) and Eq. (\ref{EOScondition}), with
five unknown functions of $r$, i.e., $\rho(r)$, $p_r(r)$,
$p_t(r)$, $\Phi(r)$ and $m(r)$. We shall adopt the approach by
considering a specific choice for a physically reasonable mass
function $m(r)$, thus closing the system.

This interior solution is now matched to an exterior Schwarzschild
solution at a junction interface, $a$. The surface stresses of the
thin shell are given by
\begin{eqnarray}
\sigma&=&-\frac{1}{4\pi a} \left(\sqrt{1-\frac{2M}{a}+\dot{a}^2}-
\sqrt{1-\frac{2m}{a}+\dot{a}^2} \, \right)
    \label{surfenergy}   ,\\
{\cal P}&=&\frac{1}{8\pi a} \Bigg(\frac{1-\frac{M}{a}
+\dot{a}^2+a\ddot{a}}{\sqrt{1-\frac{2M}{a}+\dot{a}^2}}
      - \frac{1+\omega
m'-\frac{m}{a}+\dot{a}^2+a\ddot{a}+\frac{\dot{a}^2m'(1+\omega)}{1-2m/a}}
{\sqrt{1-\frac{2m}{a}+\dot{a}^2}} \, \Bigg)
    \label{surfpressure}    \,,
\end{eqnarray}
where the overdot denotes a derivative with respect to proper
time, $\tau$. $\sigma$ and ${\cal P}$ are the surface energy
density and the tangential pressure,
respectively~\cite{darkstar2,LoboCrawford}. The dynamical
stability of the transition layer of these dark energy stars to
linearized spherically symmetric radial perturbations about static
equilibrium solutions was further explored using Eqs.
(\ref{surfenergy})-(\ref{surfpressure}) (see ~\cite{darkstar2} for
details). Large stability regions were found that exist
sufficiently close to where the event horizon is expected to form,
so that it would be difficult to distinguish the exterior geometry
of these dark energy stars from astrophysical black holes.

Several relativistic dark energy stellar configurations may be
analyzed by imposing specific choices for the mass
function~\cite{darkstar2}. For instance, consider the following
mass function
\begin{equation}
m(r)=\frac{b_0 r^3}{2(1+2b_0 r^2)}\,,
      \label{TMWmass}
\end{equation}
where $b_0$ is a non-negative constant. The latter may be
determined from the regularity conditions and the finite character
of the energy density at the origin $r=0$, and is given by
$b_0=8\pi \rho_c/3$, where $\rho_c$ is the energy density at
$r=0$. This choice of the mass function represents a monotonic
decreasing energy density in the star interior. Now the function
$\Phi(r)$ may be deduced from Eq. (\ref{EOScondition}), and is
given by
\begin{equation}
\Phi(r)=\frac{1}{2}\ln\left[\left(1+b_0r^2\right)^{(1-\omega)/2}
\left(1+2b_0r^2\right)^{\omega}\right]\,.
\end{equation}
It can be shown that the gravity profile is negative,
$\Phi'(r)<0$, for $\omega<-(1+2b_0r^2)/(3+2b_0r^2)$, indicating an
outwardly gravitational repulsion~\cite{darkstar2}, which is a
fundamental ingredient for gravastar models.

In concluding, a generalization of the gravastar picture was
explored, by considering a matching of an interior solution
governed by the dark energy equation of state, $\omega\equiv p/
\rho<-1/3$, to an exterior Schwarzschild vacuum solution at a
junction interface (An analogous case was analyzed in the phantom
regime $\omega<-1$, in the context of wormhole
physics~\cite{phantomWH}). Despite of the repulsive nature of the
interior solution, large stability regions exist, with the
transition layer placed sufficiently close to where the event
horizon is supposed to have formed~\cite{darkstar2}. It has also
been argued that there is no way of distinguishing a Schwarzschild
black hole from a gravastar, or a dark energy star, from
observational data \cite{AKL}. The dark energy stars outlined in
this paper may possibly have an origin in a density fluctuation in
the cosmological background, possibly resulting in the nucleation
of a dark energy star through a density perturbation. We would
also like to state our agnostic position relatively to the
existence of dark energy stars~\cite{CFV}, however, it is
important to understand their general properties to further
understand the observational data of astrophysical black holes.



\end{document}